\documentclass[usenatbib]{mn2e}

\usepackage{graphics}
\usepackage{epsfig}
\usepackage{natbib}

\begin{document}

\title[Physical origin of the large-scale conformity in the specific star formation rates of galaxies]
{Physical origin of the large-scale conformity in the specific star formation rates of galaxies}  

\author [G.Kauffmann] {Guinevere Kauffmann \\
\thanks{E-mail: gamk@mpa-garching.mpg.de}
Max-Planck Institut f\"{u}r Astrophysik, 85741 Garching, Germany\\}

\maketitle

\begin{abstract} 
Two explanations have been put forward to explain the observed conformity between the colours and
specific star formation rates (SFR/$M_*$) of galaxies on large scales: 1) the formation times of
their surrounding  dark matter halos are
correlated (commonly referred to as ``assembly bias''), 
2) gas is heated over large scales at early times, 
leading to coherent modulation of cooling and star formation between well-separated galaxies 
(commonly referred to as ``pre-heating'') .                               
To distinguish between the pre-heating and assembly bias scenarios,
we search for relics of energetic feedback events in the
neighbourhood of central galaxies with different specific star formation rates.  
We find a significant excess of very high mass ($\log M_* > 11.3$)  galaxies  out to a distance of 2.5
Mpc around low SFR/$M_*$ central galaxies compared to control 
samples of higher SFR/$M_*$ central galaxies with the 
same stellar mass and redshift. We also find
that very massive galaxies in the neighbourhood of low  SFR/$M_*$ galaxies 
have much higher probability of hosting radio loud active galactic nuclei.   
The radio-loud AGN fraction in neighbours with $\log M_* > 11.3$ is four times higher
around passive, non star-forming centrals at projected distances of 1 Mpc and two times higher
at projected distances of 4 Mpc.
Finally, we carry out an investigation of conformity effects in
the recently publicly-released Illustris cosmological hydrodynamical
simulation, which includes energetic input both from
quasars and from radio mode accretion onto black holes. We do not find conformity effects of
comparable amplitude on large scales in the simulations and we propose that gas needs to be
pushed out of dark matter halos more efficiently at high redshifts.
\end{abstract}

\section {Introduction} The subject of this paper is the physical origin of the
very large-scale conformity in the colours and specific star formation rates
of isolated central galaxies and their neighbours on scales in excess of 1 Mpc.

The term conformity was first introduced by Weinmann et al (2006) to describe
the conformity between the colours of central and satellite galaxies within a
single virialized group or dark matter halo. As noted in that paper, there are
obvious physical explanations for such effects if they are seen
internal to the halo virial radius. In massive dark matter halos, infalling gas
is shock heated to temperatures in excess of $10^6$ K and then cools over timescales of
of many gigayears. Cold gas in satellite galaxies moving through a halo can  
be removed  by ram-pressure effects. The combination of inefficient cooling onto
central galaxies, and stripping of gas from satellites could easily produce a
conformity effect.

An analysis showing that conformity effects extend well beyond the virial radius
for {\em isolated} central galaxies was first presented in a paper by Kauffmann
et al (2013; hereafter K13). In particular, a correlation between the specific
star formation rates of red central galaxies with stellar masses less than a few
$\times 10^{10} M_{\odot}$ and the specific star formation rates of their
neighbours was found to extend out to scales in excess of 3-4 Mpc. The strength
of the effect on large scales was surprisingly large, particularly for low mass
galaxies with little ongoing star formation.  \footnote {As seen in Figure 2 of
the paper, the  median value of SFR/M$_*$ for neighbours of galaxies in the
lowest quartile bin in  SFR/M$_*$   is depressed by more than a factor of 2
below that measured for neighbours of galaxies in the upper two quartiles, out
to distances of at least 3 Mpc.}

Explaining how correlations in star formation extend out to such large scales is
difficult, because this implies some form of causal connection between the
history of galaxies in different dark matter halos. This contradicts  the
assumption that is frequently made in the very simplest halo occupation
distribution (HOD) models, namely that the physical properties of a galaxy are
determined solely by the mass of the halo in which it resides.

Analysis of N-body simulations has shown that the amplitude of the two-point
correlation function of dark matter haloes with masses less than $10^{13}
M_{\odot}$  on large scales depends on halo formation time, an effect commonly
dubbed ``assembly bias'' (Gao et al 2005) .  If the star formation histories of
galaxies are correlated with the halo formation time, this could  be a possible
explanation for the observed large-scale conformity seen in the observations.
For this reason, Kauffmann et al (2013) undertook a direct comparison of their
observational results with the predictions of the semi-analytic models of Guo et
al (2011), which are grafted onto the exact same simulations which were used in
the  Gao et al analysis of assembly bias. The main conclusion from the
comparison  was that low amplitude conformity effects were indeed present, but
that the amplitude could not match the observational data.  Kauffmann et al
(2013) then went on to speculate  that more exotic effects, e.g. pre-heating of
the gas by energy input from active galactic nuclei, might be needed to explain
their data.

Since then, a number of theoretical papers have appeared claiming much stronger
effects arising from assembly bias alone (Wetzel et al 2014; Hearin et al 2015a,b).
The main difference between these models and those of Guo et al (2011) lies in  the
modelling of supernova feedback effects. In the Guo et al (2011) models, the
amount of cold disc gas that is reheated by SN feedback and injected into the
hot halo component scales as the fifth power of the circular velocity of the
halo. In effect, this implies that the correlation between galaxy star formation
history and  halo mass will be much stronger than the correlation with halo
formation time. Strong scaling in supernova feedback efficiency with halo mass
is required to match a number of other observables, such as the slope of the
low mass end of the stellar mass function and slope of the stellar mass-metallicity
relation at z=0. The models of Wetzel et al and Hearin et al do not take into
account these additional constraints. 

In this paper, we will try to make further progress by analyzing the
observed conformity effect in more detail. To distinguish between the
pre-heating and the assembly bias scenarios, we propose the following test.  We
ask whether we  can find any evidence for ``relics'' of present or past energetic
feedback events  in the vicinity, but outside the host halos of
low SFR/$M_*$ galaxies.  Such relics should not be detected in the
vicinity of higher SFR/$M_*$  galaxies of the same stellar mass and
redshift. We  will show  that relics in the form of an excess of
very high mass ($\log M_* > 11.3$)  galaxies are found out to a projected distance of 2.5
Mpc from passive low mass galaxies.  A significantly larger fraction of these
relics {\em currently} host radio-loud active galactic nuclei.
Finally, we carry out an investigation of conformity effects in
the recently publicly-released Illustris cosmological hydrodynamical
simulation (Vogelsberger et al 2014), which include energetic input both from
quasars and radio mode accretion onto black holes.
A Hubble constant $H_0=70$ km s$^{-1}$ is adopted
throughout the paper.

\section {New constraints from  Sloan Digital Sky Survey data}

We begin with the same volume-limited sample of 11,673  galaxies with $\log M_∗
> 9.25$ and redshifts in the range $0.017 < z < 0.03$ as in K13. The upper limit
in redshift ensures that we are able to detect all galaxies down to a limiting
stellar mass of $2 \times 10^9 M_{\odot}$, irrespective of the intrinsic colour
of the system. Galaxies with mass $M_∗$ are defined to be  central galaxies if
there is no other galaxy with with stellar mass greater than $M_∗$/2 within a
projected radius of 500 kpc and with velocity difference less than 500 km
s$^{-1}$. There are 7712 galaxies in our catalogue with stellar masses greater
than $5 \times 10^9 M_{\odot}$ to which we can apply this criterion.

For the current analysis, we select a set of passive sub-L$_*$ central galaxies
in the stellar mass range $9.75 < \log M_* < 10.25$ (i.e. similar in mass to the
Large Magellanic Cloud) that have SFR/M$_*$ values in the lowest quartile of 
the distribution. The
star formation rates we use are the {\em global} SFR values derived using the
methodology outlined in  Brinchmann et al (2004).  The stellar masses are drawn
from the MPA/JHU Data Release 7 value-added catalogue 
(http://www.mpa-garching.mpg.de/SDSS/DR7/).  For each
of these galaxies, we select four control galaxies closely matched in
stellar mass and redshift, but with SFR/M$_*$ in the upper two quartiles.
These galaxies form our control sample. In
Figures 1-4 in this section, results for the passive galaxies are shown in red,
while results for the control galaxies are shown in black.

The top two panels of Figure 1 show the cumulative total stellar mass in
neighbours as a function of projected radius around central galaxies in two
stellar mass ranges, $9.75<\log M_*<10$ and $10<\log M_*<10.25$.  All neighbours
are required to have projected velocity differences less than 500 km s$^{-1}$
from the systemic velocity of the central galaxy.  The effect of our isolation
cut to define central objects is seen as a break in the slope of the cumulative
mass profile at $R=0.5$ Mpc. Here, and everywhere else in the paper, error bars
are computed by bootstrap resampling the central galaxy samples. 
The cumulative stellar mass in neighbours  around
passive centrals is significantly higher than around  the control sample, but note that the
difference is less than 0.2 dex in $\log M_*$ at all projected radii.  The
bottom two panels show the cumulative mass in black holes in the neighbours.  We
estimate black hole mass from the measured velocity dispersion of the galaxy
using the formula in Tremaine et al (2002) $\log(M_{BH}/M_{\odot})= \alpha
+\beta(\sigma/\sigma_0)$, with $\alpha=8.13$ and $\beta=4.02$.  \footnote {The
wavelength resolution of the SDSS spectrograph means that velocity disperions
below 70 km/s cannot be reliably determined.  In our analysis we simply set the
black hole mass to zero in galaxies with $\sigma <$ 70 km/s} As can be seen, the
cumulative mass in black holes in the vicinity of the passive centrals is the
same as that around the control galaxies.

\begin{figure}
\includegraphics[width=88mm]{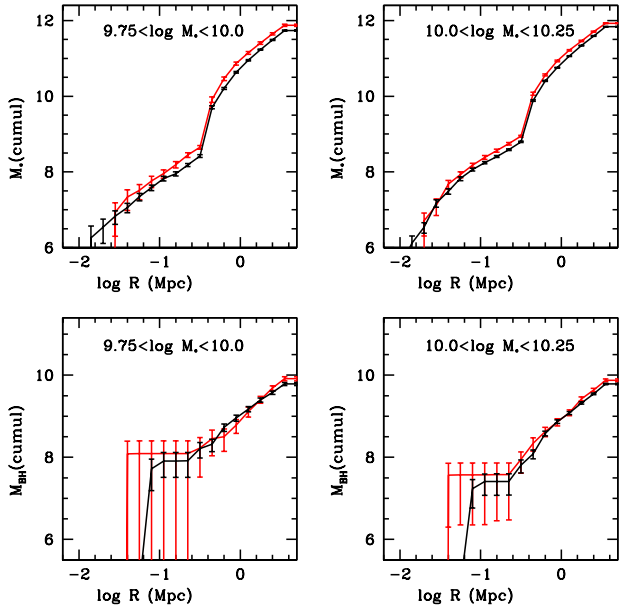}
\caption{ {\bf Top:} The cumulative total stellar mass in
neighbours as a function of projected radius around central galaxies in two
stellar mass ranges, $9.75<\log M_*<10$ and $10<\log M_*<10.25$.
The red curves show results for passive centrals; the black curves show
results for the control sample of central galaxies.
{\bf Bottom:} The cumulative total black hole mass  
as a function of projected radius around central galaxies.
\label{models}}
\end{figure}

Next, Figure 2 shows the distribution of the stellar masses of neighbours in
three radial bins centred at R=630 kpc, R=1.25 Mpc and R=2.5 Mpc. Each bin has width
0.5 in $\Delta \log R$. Note that all
three of these radial bins lie outside a projected radius of 500 kpc, i.e. the outer
radius used for the isolation criterion that defines our central galaxy sample.
We see that red central galaxies are more likely to have a very massive
neighbour with a stellar mass greater than $2-3 \times 10^{11} M_{\odot}$ than the
control central galaxies.  This excess is strongest in the closest (R=630 kpc)
bin, but remains significant out beyond 1.25 Mpc.

\begin{figure}
\includegraphics[width=88mm]{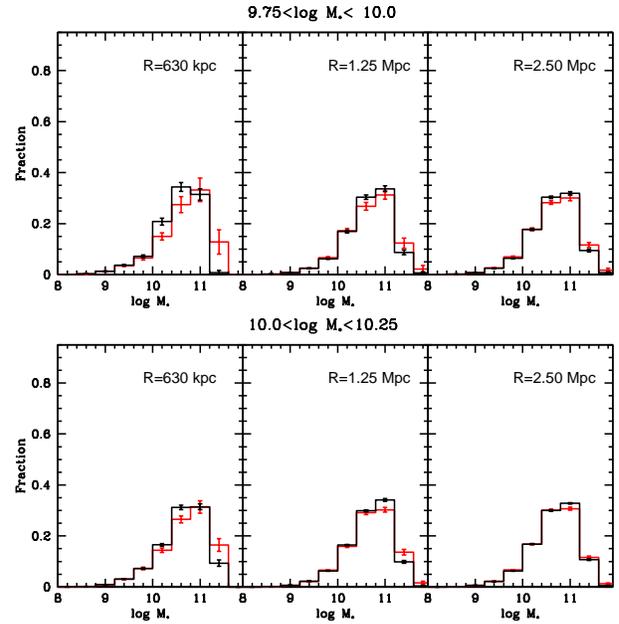}
\caption{The distribution of the stellar masses of neighbours in
three radial bins centred at R=630 kpc, R=1.25 Mpc and R=2.5 Mpc.
Results are shown for the stellar mass ranges, $9.75<\log M_*<10$ (top)
and $10<\log M_*<10.25$ (bottom).
The red histograms show results for passive centrals; the black histograms show
results for the control sample of centrals.
\label{models}}
\end{figure}

We now investigate whether peculiarities exist in the {\em structural}
properties of the neighbours in the vicinity of passive galaxies with low
stellar masses. To do this, we convert the $g$ and $i$-band surface brightness
profiles of each neighbouring galaxy into a stellar mass density profile.  The
procedure for doing this is described in Kauffmann et al (2007). The SDSS {\em
frames} pipeline extracts an azimuthally-averaged radial surface brightness
profile for each galaxy. In the catalogs, this is  given as the average
surface brightness in a series of annuli, from which interpolated colour and
stellar mass density profiles can be derived.  The conversion between $g-i$
colour and $i$-band mass-to-light ratio is given by the relation shown in Figure
15 of Kauffmann et al (2007).

In Figure 3, we show how the total stellar mass in in neighbours around passive
central galaxies and the control sample is partitioned as a function of {\em local} stellar
mass surface density.  Results are shown for the same three radial bins as in
Figure 2. No significant difference is seen in the stellar mass density
distrbution functions for passive centrals and control galaxies .  We conclude
that the main distinguishing feature of the population of galaxies in the
neighbourhood of passive, low mass centrals, is the more probable  presence of a
giant galaxy with $\log M_* > 11.3$.

\begin{figure}
\includegraphics[width=88mm]{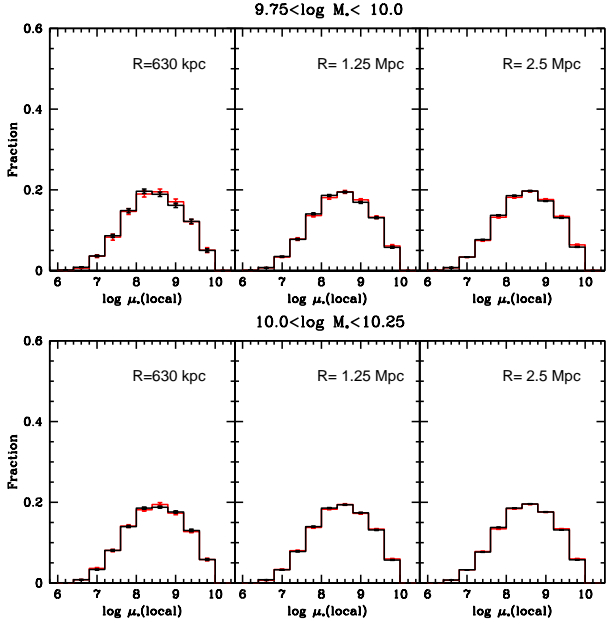}
\caption{ The distribution of the  total stellar mass in neighbours 
as a function of local stellar
mass surface density. Results are shown for the same three radial bins
and the same stellar mass bins  as in Figure 2.
\label{models}}
\end{figure}

\subsection {Radio-loud AGN fraction} Best et al (2005) showed that the fraction
of galaxies that host radio-loud AGN is a strong function of stellar mass,
rising from values close to zero at $10^{10} M_{\odot}$ to 30\% at $M_* = 5
\times 10^{11} M_{\odot}$.  In follow up work, Best et al (2007) also  showed
that at fixed mass, central galaxies in groups and clusters had higher
probability of being radio loud than galaxies of the same mass selected
independent of their environment. The difference in radio-loud fraction was
largest for low mass galaxies (a factor 10 at $M_* \sim 3 \times 10^{10}
M_{\odot}$), but negligible for the highest mass galaxies (almost no difference
for galaxies with masses greater than $3 \times 10^{11} M_{\odot}$).

In Figure 4, we plot the cumulative fraction of neighbouring  galaxies interior to 
a projected radius R that host radio-loud AGN.
Results for neighbours with $\log M_* > 11$
are shown as dotted lines, and for neighbours with $\log M_* > 11.3$ are shown as
solid lines. Red lines are for passive centrals and black lines are for the
control sample. Radio AGN are identified using the most recent release of a
catalogue of 18,286  systems constructed by combining the seventh data release
of the Sloan Digital Sky Survey with the NRAO (National Radio Astronomy
Observatory) VLA (Very Large Array) Sky Survey (NVSS) and the Faint Images of
the Radio Sky at Twenty centimetres (FIRST) survey (Best \& Heckman 2012).

\begin{figure}
\includegraphics[width=78mm]{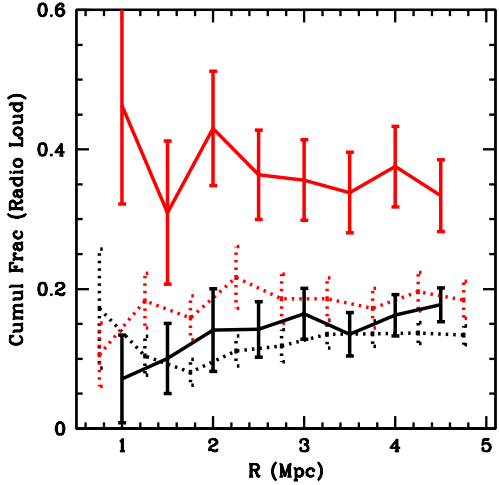}
\caption{ The cumulative fraction of neighbouring  galaxies interior to radius R
that host radio-loud AGN.
Results for neighbours with $\log M_* > 11$
are shown as dotted lines, and for neighbours with $\log M_* > 11.3$ are shown as
solid lines. Red lines are for passive centrals and black lines are for the
control sample.  
\label{models}}
\end{figure}

As can be seen, the fraction of radio-loud AGN hosted by massive neighbours is
significantly higher aound the passive centrals than it is around the control
sample. In contrast to the trends found  for  central galaxies by Best et al (2007), the
{\em difference is largest for the most massive neighbours}. The excess in radio-loud
AGN fraction also  appears to extend out to very large radii.
Among neighbours with $\log M_* > 11.3$, the
difference in radio-loud fraction between the passive centrals and the controls
is a factor of 4 at R=1 Mpc and is still very significant  (a factor of 2) at R=4 Mpc.

We conclude that passive low mass central galaxies have high probability of marking 
the location of patches of the Universe extending over many Mpc
where there is a clear excess of massive galaxies hosting radio-loud AGN.
Figures 5 and 6 show six examples of such massive relics located within
a projected radius of  5 Mpc from  a
low SFR/$M_*$ central galaxy and with velocity difference $\Delta V <$ 500 km/s. 
Optical images from the SDSS Skyserver are displayed in
Figure 5 and radio continuum cut-out images from the FIRST/VLA Image Cutout server
are displayed in Figure 6. The optical images show that the massive galaxies are
all normal giant ellipticals, with red colours and featureless morphologies.
The radio morphologies of the AGN are quite varied -- some are compact and unresolved,
while others  extend over several arcminutes in diameter and exhibit clear double-lobed
structures. The most extended of the sources (middle bottom panel) has a
physical diameter of 70 kpc. In this case, the radio jets do extend outside the optical
radius of the host galaxy. Nevertheless, it is clear from these
images that these these jets are not currently
dumping energy over scales of many Megaparsecs. We speculate instead that
much of the gas in the halos of these systems was heated and expelled at early epochs during the formation
of the massive galaxy itself, and that the current radio-loud AGN activity simply
acts to prevent gas from re-cooling, and to maintain the galaxy in a passive state, similar
to what is seen in present-day clusters (see Fabian (2012) for a review).

\begin{figure}
\includegraphics[width=95mm]{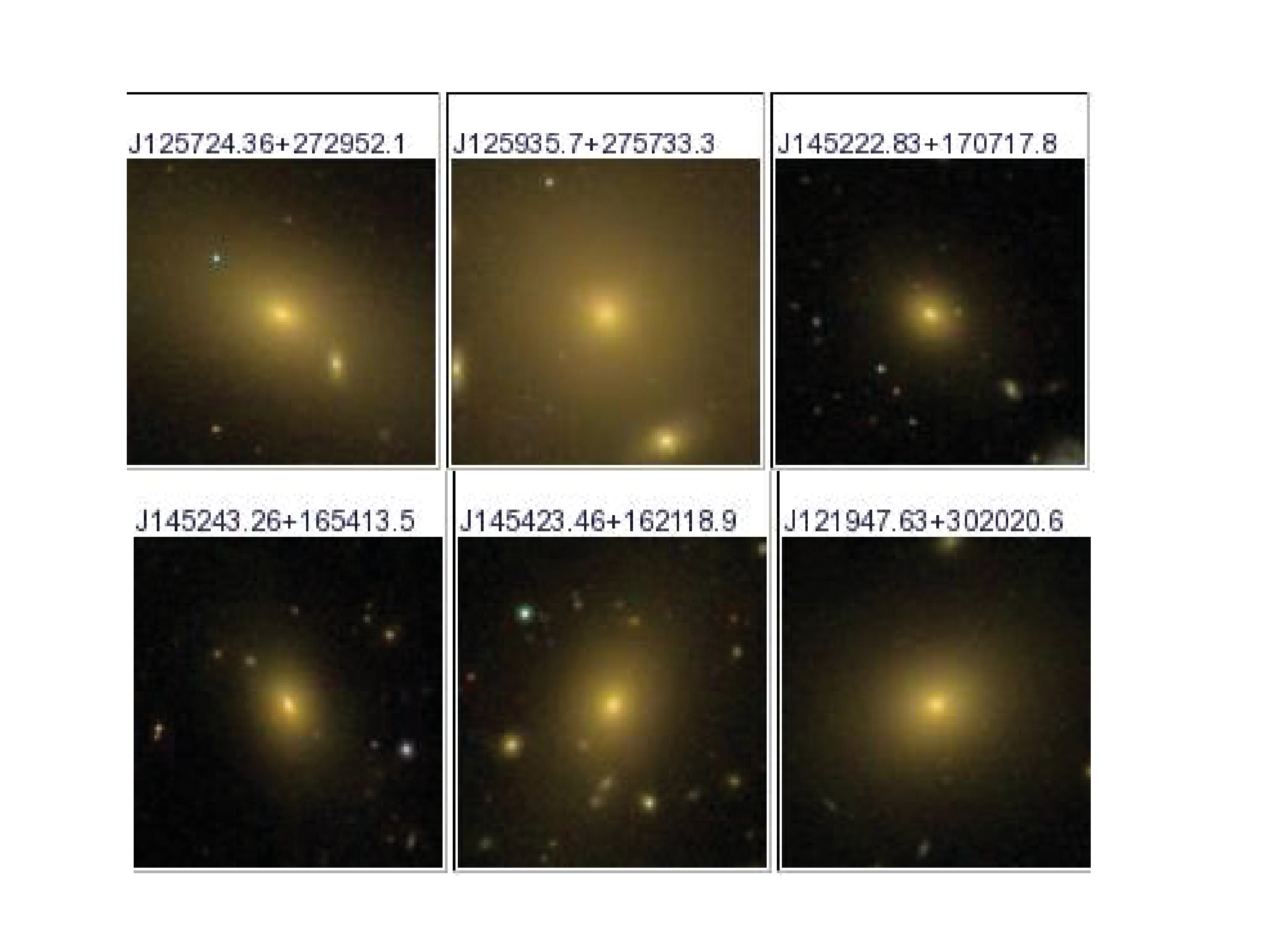}
\caption{ Optical images of six examples of very massive ($\log M_* > 11.3$) galaxies that host
radio-loud AGN and that are found within a projected radius $R<5$ Mpc 
from a passive low mass central galaxy  
with velocity difference $\Delta V <$ 500 km/s.
\label{models}}
\end{figure}

\begin{figure*}
\includegraphics[width=120mm]{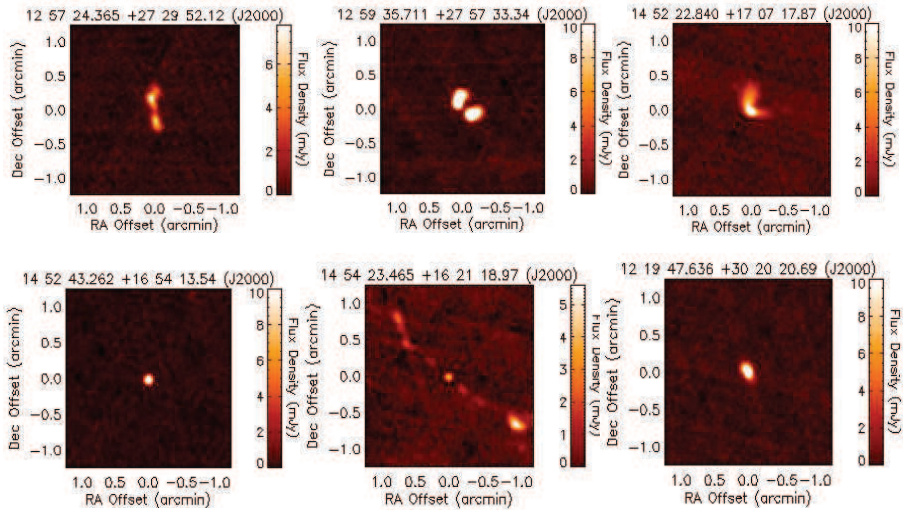}
\caption{ 1.4 GHz radio continuum images of six examples of very massive ($\log M_* > 11.3$) galaxies that host
radio-loud AGN and that are found within a projected radius $R<5$ Mpc 
from a passive low mass central galaxy  
with  velocity difference $\Delta V <$ 500 km/s.
\label{models}}
\end{figure*}

\section {Comparison with Illustris simulation} As discussed in Section 1,
Kauffmann et al (2013) compared their observational results with the
semi-analytic models of Guo et al (2011). They selected central galaxies and
neighbours in the same way as in the observations and found that conformity
effects on large scales were too weak to match observations.  We note that there
is no ``cross-talk'' between gas-physical processes in different dark matter
halos in these models. Even though supernovae may eject gas out of the dark
matter halo, this gas is simply held in limbo in a so-called ejected reservoir
which will eventually re-join the same dark matter halo after a fixed delay. In
more recent work, Henriques et al (2013) propose that these re-incorporation times
need to be longer in lower mass halos in order to match observed constraints on
the evolution of the galaxy mass function at low stellar masses. In these
circumstances, the assumption that the baryon reservoir of a halo is always tied
to that halo over all cosmic time, is even more contrived.

In numerical hydrodynamical simulations, the thermodynamic
evolution of the gas is followed in a self-consistent way. Recently, the
Illustris simulation project has publicly released  a suite of large volume,
cosmological hydrodynamical simulations run with the moving-mesh code Arepo
(Nelson et al 2015). The simulations include a complex set of prescriptions for
following a whole set of physical processes relevant for the formation and
evolution of galaxies (Vogelsberger et al. 2014). In particular, detailed
prescriptions for energetic feedback from black holes by quasars triggered
during galaxy-galaxy mergers and by bubbles of hot gas produced by radio-loud
AGN when black holes accrete at low rates (Sijacki et al 2007; 2015) are included
in the simulation code.  The Illustris simulation has been demonstrated to
provide  much better match to basic galaxy properties such as the stellar mass
function and stellar mass/halo mass relation than was typical of previous
generations of simulations. It thus serves as a useful
benchmark for comparisons with a wide variety of different observational data.

In this section, we will carry out two direct comparisons between our SDSS data
and Illustris: 1) We will select a set of isolated central galaxies from the
simulation using the same procedure that we use for the observations, and we will ask
whether the distribution of specific star formation rates matches what is seen
in the observational data. The reason for carrying out the comparison using
central galaxies rather than all galaxies, is to separate the effects of
internal processes that regulate star formation in isolated  galaxies (i.e.
supernova and AGN feedback) from the effects of evironmental processes such as
ram-pressure stripping of gas in group and cluster environments.  2) We examine
whether conformity between the colours of the central galaxies and those of
their neighbours is seen at separations larger than 1 Mpc.

\subsection {Specific star formation rate distributions of isolated central
galaxies} We extract $x,y,z$ positions, velocities, stellar masses and star formation
rates for galaxies in the z=0  SUBFIND subhalo catalog (see Nelson et al. (2015) for
more details). We adopt stellar masses and star formation rates measured within
twice the half mass radius to derive our fiducial measures of SFR/$M_*$  and we consider only those
subhalos containing galaxies with stellar masses greater than $2 \times 10^9 M_{\odot}$. We select
isolated central galaxies the same way as in the data, by requiring that no
other galaxy more massive than half the mass of the object be found within a
radius $dr=\sqrt {dx^2+dy^2}$ less than 500 kpc and within a velocity difference
$H_0dz +(dv_z)$ less than 500 km/s.

Figure 7 shows histograms of $SFR/M_*$ for galaxies in three different bins of
stellar mass: $9.75<\log M_*<10$, $10.75<\log M_*<11$, and $11.25<\log M_*<11.5$.
The upper panels show results for the SDSS, while the lower panels show results
for Illustris. In order to illustrate the systematic uncertainties arising from
the aperture within which we measure $SFR/M_*$, we plot distributions for
$SFR/M_*$ measured within the 3 arcsec fibre aperture as dotted histograms in
the upper panels. In the lower panels, the dotted histograms are for $SFR/M_*$
measured within the radius containing half the stellar mass. Independent of
which measure of $SFR/M_*$ is adopted, significant discrepancies between the
simulations and the data are apparent: \begin{enumerate} \item In the lowest
mass bin, there is a significant tail of galaxies with $\log SFR/M_* < -11$,
which is missing in the Illustris simulation. The $SFR/M_*$ values of the
simulated galaxies are all peaked in a very narrow range around $\log SFR/M*
\sim -10$, which implies that these galaxies have formed stars at a constant
rate over the age of the Universe.  \item In the bin with $10.75<\log M_*<11$,
we see that the bulk of the population in the SDSS have low $SFR/M_*$, whereas
the opposite is true in Illustris. \item In the highest mass bin, essentially
all galaxies in SDSS are passive with $\log SFR/M_* < -11$, but Illustris has a
much larger fraction of galaxies which are still actively forming stars.
\end {enumerate}  
We note that Sparre et al (2015) carried out a comparison of only the star-forming
``main sequence'' with observational data at different redshifts, and did not
note the discrepancy in passive galaxy fractions as a function of stellar mass
at $z=0$.

Our main conclusion, therefore, is that star formation quenching processes
appear to work too inefficiently in Illustris compared to the real Universe.  We
will speculate on reasons why this might be the case in the final section.

\begin{figure}
\includegraphics[width=88mm]{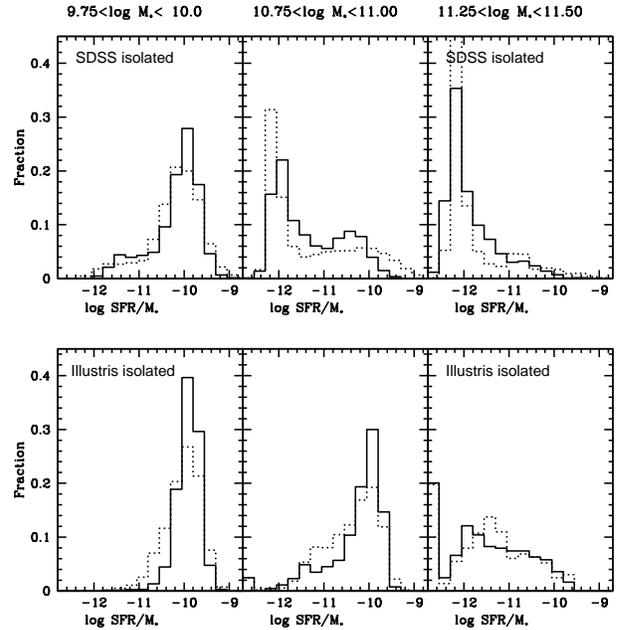}
\caption{ Histograms of $SFR/M_*$ for central isolated galaxies in three different bins of
stellar mass: $9.75<\log M_*<10$, $10.75<\log M_*<11$, and $11.25<\log M_*<11.5$.  
The upper panels show results for the SDSS, while the lower panels show results
for Illustris. In the top panels, the solid histograms are for the global SFR/$M_*$
measurements from the MPA/JHU catalogue (see Brinchmann et al. (2004) for details), while
dotted histograms are for SFR/$M_*$ measured within the 3 arcsec SDSS fibre.  
In the bottom panels, the solid histograms are for  SFR/$M_*$ measured within
twice the half mass radius of the main galaxy of the subhalo
(see Nelson et al. (2015) for details), while
dotted histograms are for SFR/$M_*$ measured within the half mass radius. 
\label{models}}
\end{figure}

\subsection {Evidence for conformity?} Given that the tail of passive low mass
galaxies seen in SDSS data is not present in the Illustris simulation and that high
mass galaxies form stars at rates that are too high compared
with observations, one might question the
validity of carrying out a conformity analysis. Nevertheless, in view of the         
different predictions for the effect of assembly bias on the colours and star formation
rates of galaxies in the literature, we feel that it is useful to examine
conformity effects in a numerical hydrodynamical simulation that was not explicitly tuned
to reproduce this effect.

Figure 8 shows the specific star formation rates
of neighbouring galaxies as a function of projected radius around $3 \times 10^9
-10^{10} M_{\odot}$ central galaxies in the lowest 25th percentile bin in
$SFR/M_*$ (red lines) and in the 50th-75th percentile bin (black lines). Solid
lines show the median $SFR/M_*$ of the neighbours, while dashed and dotted lines
show the 25th and 75th percentiles of the SFR/$M_*$ distribution. The left panel
shows results for galaxies from the Sloan Digital Sky Survey. We now plot our
results over the projected radius range 1-30 Mpc and we find that the effect
appears to persist out to a radius of $\sim 10$ Mpc in this mass bin. Results
for Illustris are plotted in the right panel.

The first-order discrepancy in the {\em dynamic range} in SFR/$M_*$ spanned by
neighbouring galaxies is again very apparent.  Similar to what was found for the
Guo et al (2011) semi-analytic models, the depression in star formation rate
around passive centrals in Illustris at projected radii larger than 1 Mpc is
very small and is mainly confined to the lowest 25th percentile of the $SFR/M_*$
distribution. 

\begin{figure}
\includegraphics[width=90mm]{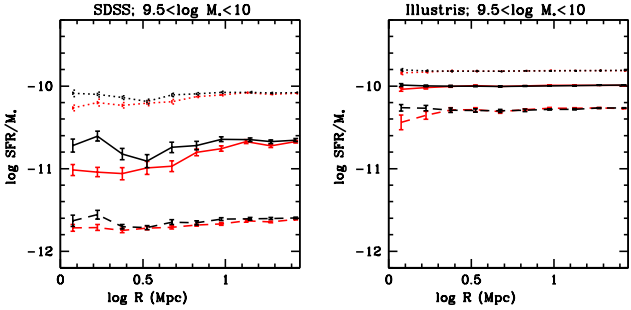}
\caption{ The specific star formation rates
of neighbouring galaxies is plotted as a function of projected radius around $3 \times 10^9
-10^{10} M_{\odot}$ central galaxies in the lowest 25th percentile bin in
$SFR/M_*$ (red lines) and in the 50th-75th percentile bin in $SFR/M_*$ (black lines). Solid
lines show the median $SFR/M_*$ of the neighbours, while dashed and dotted lines
show the 25th and 75th percentiles of the SFR/$M_*$ distribution. The left panel
shows results for galaxies from the Sloan Digital Sky Survey, while the right panel is for 
Illustris.  
\label{models}}
\end{figure}

\section {Summary and Discussion} In this paper, we have attempted to
further our understanding of the  large scale correlations between the
specific  star formation rates in
isolated low mass galaxies and their distant neighbours. In
particular, we test the hypothesis that energetic feedback from an early generation of accreting black
holes may be responsible for heating gas to high temperatures over large
scales. We find an excess  
number of very  massive galaxies out to a radius of 2.5
Mpc around passive central galaxies. In addition, we find that the probability
for these massive galaxies to host radio loud AGN at the present day is 3-4
times higher compared to massive neighbours around the control sample.
Interestingly, this radio AGN excess extends out to 
projected radii beyond 5 Mpc.  We have compared our results
to the Illustris simulation, which includes energetic feedback by both a quasar
and radio galaxy population, but do not find conformity effects of comparable
strength to those seen in the data.

Recent work by Choi et al (2015) has highlighted how different prescriptions for
AGN feedback can have profound effect on the properties of the gas around
massive ellipticals.  By switching from a thermal feedback prescription
similar to the one used in the Illustris simulation to a mechanical and
radiation feedback model that operates during high black hole accretion rate
growth phases, the X-ray luminosities of massive ellipticals are found to decrease by a
factor of 10-100, because much more of the gas is blown out beyond the halo
virial radius, providing a much better fit to the observed X-ray luminosities of
elliptical galaxies at the present day.

Bah\'e \& McCarthy (2013) have used cosmological hydrodynamical
simulations to show that direct ram pressure interaction with an extended gas
`halo' surrounding massive galaxies can be sufficiently strong to strip the
gaseous atmospheres of infalling galaxies out to distances of up to 5 virial
radii.  We might also even speculate that the extended hot gas left over from an
early violent gas ejection event forms  a reservoir that is able to fuel black
holes at low rates and trigger radio-mode AGN at later epochs.  These radio-loud
AGN in turn prevent gas from cooling efficiently, leading to large, coherent
zones of the Universe where galaxies can no longer form any stars.

Recent observational evidence that gas has been expelled from a substantial
number of massive studies has come from studies of the X-ray emission from from
˜250 000 ``locally brightest galaxies'' selected from the Sloan Digital Sky Survey
by Anderson et al (2015). These authors apply the same cut employed in this
paper to select central galaxies and then stack X-ray data from the  ROSAT
All-Sky Survey to recover the mean X-ray luminosity as a function of galaxy mass
down to $M_*=10^{10.8} M_{\odot}$. Interestingly, the relation they infer between
X-ray luminosity and dark matter halo mass has normalization more than a factor
of 2 below relations that have been derived based on X-ray-selected cluster
samples. This suggests that there is a lot of intrinsic scatter between X-ray
luminosity and halo mass. The authors speculate that one explanation for this
scatter is that gas has been pushed outwards in the low luminosity systems.
High sensitivity maps of hot gas in wide range of different environments using
next-generation facilities such as the  eROSITA telescope will be needed to make
further progress on this topic. It is currently necessary to stack thousands of
galaxies in ROSAT to recover a signal, but eROSITA should enable similar
analyses  for individual galaxies, or for smaller stacked samples, allowing us
to pinpoint the conditions under which gas is confined within dark matter halos
and the conditions that lead to its expulsion.

It remains to be seen whether changes to AGN feedback prescriptions in Illustris will be sufficient to
bring the specific star formation rate distributions of simulated galaxies into better
agreement with observations. Sparre et al (2015) demonstrate
that there are too few strongly starbursting galaxies in the simulation and
propose that the fixed, stiff equation of state used to describe the
interstellar medium may require modification. The fact that virtually all low
mass glaxies in the simulation have $SFR/M_* \sim -10$ suggests that the
interstellar medium does not respond very sensitively to the  dynamical history of
the galaxy. In contradiction with the the written claims in their paper, Figure 1 of
Sales et al (2015) reveals rather poor agreement between the density profiles of blue and
red satellite galaxies in Illustris and SDSS. In particular, the differences
between the profiles for blue and red galaxies are larger for low mass galaxies
in SDSS, but the {\em opposite} is seen in the simulations. Further comparisons
with the EAGLE simulations, which have similar resolution and claim even better
agreement with the stellar mass function and the $M_*-M_{halo}$ relation (Schaye
et al 2015), but which include very different implementation of much of the
galaxy formation physics, will be useful for generating improvements to the next
generation of simulations.

\section*{Acknowledgments}
I thank Aaron Bray for helpful discussions.
Thank you to the Aspen Center for Physics and the NSF
Grant No. 1066293 for hospitality and supporting 
minds during the analysis phase and writing of this paper.


\end{document}